\documentstyle[prl,aps,preprint,tighten,floats,epsf]{revtex}

\begin{document}
%\jl{1}

\newcommand{\fig}[2]{\epsfxsize=#1\epsfbox{#2}}

\title{Stability of the Bragg glass phase in a layered geometry}
\author{David Carpentier and Pierre Le Doussal}
\address{CNRS-Laboratoire de Physique Th\'eorique de l'Ecole\\
Normale Sup\'erieure, 24 rue Lhomond, F-75231 Paris\cite{frad}}
\author{Thierry Giamarchi}
\address{Laboratoire de Physique des Solides, Universit{\'e} Paris-Sud,\\
B{\^a}t. 510, 91405 Orsay, France\cite{junk}}
\date{\today}
\maketitle

\begin{abstract}

We study the stability of the dislocation-free
Bragg glass phase in a layered geometry consisting of
coupled parallel planes of d=1+1 vortex lines
lying within each plane, in the presence of impurity
disorder. Using renormalization group, replica variational
calculations and physical arguments
we show that at temperatures $T<T_G$
the 3D Bragg glass phase is always stable for weak 
disorder. It undergoes
a weakly first order transition into a decoupled 2D vortex
glass upon increase of disorder. 
\end{abstract}

An important problem is to
understand how impurity disorder
affects the translational long range order of periodic
media such as the Abrikosov vortex lattice \cite{blatter_vortex_review},
charge density waves (CDW) \cite{gruner_revue_cdw},
Wigner crystals \cite{wigner_andrei} and
magnetic bubbles \cite{seshadri_bubbles_long}.
A purely elastic theory starting from the crystal,
assuming the absence of topological defects,
such as dislocations, predicts an algebraic decay
of translational order
\cite{nattermann_vortex,giamarchi_vortex_global,%
korshunov_variational_short,villain_cosine_realrg} and that the resulting
glass phase has divergent Bragg peaks
\cite{giamarchi_vortex_global}. This is in marked
contrast with other models of glasses in superconductors,
such as the ``vortex glass'' \cite{fisher_vortexglass_long} or
``hexatic glass'' \cite{chudnovsky,grier_decoration} 
where dislocations were argued to be present due to disorder beyond the
Larkin length and translational order
decays rapidly. It was therefore
claimed \cite{fisher_vortexglass_long}
that the results of the elastic theory are unstable
to topological defects.
However, in \cite{giamarchi_vortex_global} it was
shown, using energy arguments, that as a consequence
of the slow decay of correlations
unbounded dislocations should not be present at equilibrium
and at weak disorder. 
The resulting phase, called ``Bragg glass'', is a
glass with topological order.
It has two important length scales, the Larkin-Ovchinnikov
length $R_c$ which characterizes pinning, and the 
translational order correlation length $R_a$. 
$R_a$ can be much larger than $R_c$ at low temperature if the scale $r_f$ 
at which disorder varies (i.e the superconducting coherence length $\xi$) is
small compared to the lattice spacing $a$. Since $R_c$ is renormalized
upwards by thermal fluctuations, one expects
these two lengths to be of the same order of magnitude
at higher temperatures and near melting.
Such a phase is compatible with present decoration and neutron
experiments \cite{grier_decoration,yaron_neutron,giamarchi_comment-neutrons}.
The random field 3D XY model
being described by a very similar model, the 
same arguments in favor of a quasi-ordered
vortex free phase apply at weak disorder.
The existence of this topologically ordered phase
in the 3D XY model
now finds some support from numerical simulations
\cite{gingras_huse}. Also, detailed scaling arguments and a variational
calculation supporting the 
existence of the Bragg glass
were constructed recently in \cite{nattermann_kierfeld}.

In \cite{giamarchi_vortex_global} the domain of validity of the elastic theory 
in presence of disorder was estimated to be $R_a \gtrsim a$. As we showed,
this implies slow decay of correlations and, using energy arguments
\cite{giamarchi_vortex_global}, the absence of dislocations. 
The Bragg glass is therefore a self-consistent solution until $R_a \gtrsim a$.
Beyond, the Bragg glass must
undergo a transition into an amorphous state
containing topological defects,
upon increase of disorder or field. The nature of this state
is unknown, various possibilities are
a strongly disordered vortex glass, an hexatic glass,
or a strongly pinned liquid.
A scenario suggested in \cite{giamarchi_vortex_global},
is that this transition occurs around the
field $H_{tr}$ of the observed tricritical point \cite{safar_tricritical_prl}
which separates, on the thermal melting line, first order melting at low field 
from second order vortex glass 
transition at higher fields. From usual 3D melting considerations, one expects
that as long as $R_c \sim R_a \gg a$ 
the weakly disordered Bragg glass should melt through a
first order transition upon raising temperature, while the amorphous state melts
through a second order transition.
One can now find increasing experimental support for 
this scenario both in YBCO and BSCCO: the tricritical point can be lowered in 
field by controlled increase of disorder \cite{safar_tricritical_prl}, 
the ''second peak'' around
$H_{tr}$ appears as a sharp separation of two different types of behaviour 
in magnetic and transport properties \cite{khaykovich_zeldov,yeshurun} and neutrons 
\cite{forgan} also indicate a sudden loss of tranlational order 
at about $H_{tr}$. 

The above self-consistency of the Bragg glass phase relies
on arguments about dislocations. Also, to study analytically the
transition between the two glass phases in $d=3$ is difficult since 
one must be able to describe the proliferation of topological defects.
An interesting model geometry to study dislocation loops in presence of disorder 
was proposed by Kierfeld et al. \cite{nattermann_kierfeld},
who also derived more detailed energy arguments in favour 
of the existence of the Bragg glass up to $R_a \sim a$.
The model consists in a model of coupled
parallel layers of $d=1+1$ lines constrained
to lie within the planes and allows to describe
some topological defects. It is relevant to the case of a magnetic field
directed along the $ab$ plane. This model was previously studied
in the absence of disorder \cite{mikheev}, and found to exhibit a transition
at $T_c$ between a 3D coupled solid for $T< T_c$ and a decoupled
high temperature 2D phase. Since the flux lines are confined
to the planes, an elastic description can still be used at weak disorder
and therefore the methods of \cite{giamarchi_vortex_global}
to describe the Bragg glass can be extended to this particular geometry.
In the present paper we give a detailed solution of this model
using both the replica variational method
and a renormalization group (RG) treatment.
We show that at weak disorder the Bragg glass phase
is stable, and that the criterion
given in \cite{giamarchi_vortex_global} properly estimates
the domain of stability of the model. Similar conclusions
were reached by Kierfeld et al.
\cite{nattermann_kierfeld} using physical argument
and a variational method.

Interacting flux lines confined to one plane
in $d=1+1$, of average spacing $a$ can be described in
the continuum limit by introducing a smooth labelling phase
field $\phi=2 \pi u/a$, where $u$ is the displacement field
in the elastic limit. Since there are no topological defects
within one plane the density of lines in each plane
can be decomposed 
\cite{giamarchi_vortex_global,nattermann_lyukutusov} as
$\rho(x)=\rho_0(1-\nabla u(x) + \sum_{p \neq 0} \exp(i p (2 \pi x+ \phi(x))))$
where $p$ are integers.
The model studied here consist of $N$ planes of vortices
coupled via their local density and in the presence of weak
impurity disorder modelled by a random potential $V(x)$ with
correlations $\overline{V_k V_{-k}}=\Delta_k$.
The model is defined by the following
random field Hamiltonian \cite{footnote1}:
\begin{eqnarray} \label{2DXY}
H &=& \int d^2x\; \sum_{i j} \frac{1}{2}
K^{-1}_{i j} \nabla \phi_{i}(x) \cdot \nabla \phi_{j}(x)
- \sum_{i} \eta_{i}(x) \cdot \nabla \phi_{i}(x) \nonumber \\
&-&
\mu_{i j} \cos(\phi_{i}(x)- \phi_{j}(x))
- \sum_{p,i} Re( \zeta^p_i e^{i p \phi_{i}(x)} )
\end{eqnarray}
with isotropic gaussian disorder
$\overline{ \zeta^p_i(x) {\zeta^p_j}^* (x') }
= 4 T g^p_{i j} \delta(x-x')$ and
$\overline{ \eta^i_{\alpha}(x) \eta^j_{\beta}(x') }
= T \Delta_{i j} \delta_{\alpha \beta} \delta(x-x')$.
The original model in its simplest isotropic version, is defined by
a smaller number of parameters:
\begin{eqnarray} \label{original}
K^{-1}_{i j} = c \delta_{i j} ~~ &;& ~~ 
\mu_{i j} = \frac{\mu}{2} ( \delta_{i+1,j}
+ \delta_{i-1,j} ) \\
g^p_{i j} = g^p \delta_{i j}  ~~& ; &~~  \Delta_{i j} = \Delta \delta_{i j}
\end{eqnarray}
$g^p=\rho_0^2 \Delta_p/T$ is proportional to the disorder in
plane $i$ with Fourier component close to $2 \pi p/a$,
$\Delta=\rho_0^2 \Delta_0/T$ is the long wavelength
disorder coupling to slow variations of
density \cite{giamarchi_vortex_global}.
The extra terms present in (\ref{2DXY}), i.e the
longer range couplings $\mu_{i j}$, interplane disorder
correlations $g^p_{i j}$, long wavelength
couplings and disorder correlators between planes, are generated
by renormalization. They must thus be added to the general
model which includes all most relevant terms
allowed by symmetry.

We start by applying the replica Gaussian
Variational Method (GVM) to the replicated
version of the starting Hamiltonian (\ref{original}):
\begin{eqnarray} \label{replicated2}
H &=& \int d^2x\; \sum_{a i} \frac{1}{2}
c (\nabla \phi^a_{i})^2
-  \mu
\cos(\phi^a_{i}(x)- \phi^a_{i+1}(x)) \nonumber \\
&& - \sum_{a b i} \frac{\Delta}{2} \nabla \phi^a_{i} .
\nabla \phi^b_{i} - g \cos(\phi^a_{i}(x)- \phi^b_{i}(x))
\end{eqnarray}
where $a=1,\dots,n$ is the replica index and one takes the limit
$n \to 0$ at the end.
For clarity have kept only the lowest harmonic $g=g^1$,
but the general case will be discussed below. Note that
for a single harmonic model $R_c \sim R_a$ \cite{giamarchi_vortex_global}.
We study $N$ coupled planes with $N \to \infty$ and
use periodic boundary conditions. All planes are then equivalent and
it is convenient to introduce the
Fourier transform along $z$,
$\phi^a_{j}(q)=\int_{q_z} \phi^a(q,q_z)
\exp[ i( j q_z)]$. We denote $\int d^2 q/(2 \pi)^2$ by $\int_q$
and $l \int_{-\pi/l}^{\pi/l} dq_z/(2 \pi)$
by $\int_{q_z}$ where $l$ is the interplane distance.

One can approximate
(\ref{replicated2}) by the variational Hamiltonian
\cite{mezard_parisi}
$H_0 = \frac{1}{2} \int_{q,q_z} G(q,q_z)^{-1}_{a b}
\phi^a(q,q_z) \phi^b (-q,-q_z)$.
The propagator $G(q,q_z)^{-1}_{a b}$ is determined
by optimizing the variational free energy
$F_{var} = F_0 + \langle H-H_0 \rangle _{H_0}$.
It is parametrized by a connected (thermal)
part and a (disorder) self-energy part
as $G^{-1}_{a b}= \delta_{ab} G_c^{-1} - \sigma_{ab} - \Delta q^2$
with $\sum_a \sigma_{a b}=0$.
The self-consistent saddle point equations
\cite{carpentier_decoupling_long} read:
\begin{eqnarray}
& & G_c(q,q_z) = 1/(c q^2 +
2 \tilde{\mu} (1 - \cos(q_z l))) \\
& & \tilde{\mu} = \mu~e^{- \frac{1}{2} B^{a a}_{i i+1}}  ~~
\sigma_{a \neq b } = 2 g~e^{- \frac{1}{2} B^{a b}_{i i}}
\nonumber \\
& & B^{a a}_{i i+1} = 2 T \int_{q,q_z} (1-\cos(q_z l)) G_{a a}(q,q_z)
\label{interplane} \\
& & B^{a b}_{i i} = 2 T \int_{q,q_z}  G_{a a}(q,q_z) - G_{a b}(q,q_z)
\end{eqnarray}
the latter being the intraplane and off-diagonal
interplane mean squared phase fluctuations, respectively.
The quantity $\tilde{\mu}$ is the effective coupling between the
planes and when it is non zero the problem is effectively 3D,
whereas $\tilde{\mu}=0$ is the signature of decoupling
and corresponds to unbound dislocations being present
between the planes.

These equations have several types of solutions,
corresponding to the three phases of the model:
high-temperature (replica symmetric),
3D coupled elastic glass with
full Replica Symmetry Breaking (RSB) and 2D decoupled
glass (one step RSB). Each of these solutions is
qualitatively similar to the ones obtained in
\cite{giamarchi_vortex_global}, for $d=3$ and $d=2$.
Here however the model naturally exhibits transitions
between these solutions.

We first study the full
RSB solution, as in \cite{giamarchi_vortex_global}.
Parametrizing $\sigma_{ab} \to
\sigma(u)$ and $B^{ab} \to B(u)$, $0<u<1$, one has
$\sigma(u)= 2 g e^{-1/2 B_{i i}(u)}$. The
inversion formula for hierarchical matrices gives:
\begin{equation} \label{hierarch}
B_{i i}(u) = B_{i i}(u_c) + \int_{q,q_z}
\int_u^{u_c} dv
\frac{\sigma'(v)}{(c q^2 +  2 \tilde{\mu} (1 - \cos(q_z l )) +  [\sigma](u) )^2}
\end{equation}
and $B_{i i}(u_c)=2 T \int_{q,q_z} (c q^2 +  2 \tilde{\mu} (1 - \cos(q_z l )))^{-1}$.
Differentiating the saddle point equation with respect to $u$,
integrating over momenta and using
$[\sigma]'=u \sigma'$ one finds
the full RSB solution for $u<u_c$:
\begin{equation}
\sigma(u) = \frac{ 2 \tilde{\mu} v/\tilde{T} }{ \sqrt{ 1 - v^2 } }
\ \ ;\ \ 
{[\sigma]}(u) = 2 \tilde{\mu} \left( \frac{1}{ \sqrt{ 1 - v^2} } - 1\right)
\end{equation}
where the variables $\tilde{T}=T/T_c$ and $v=u/\tilde{T}$ have been
introduced for convenience, $T_c=4 \pi c$ being the transition
temperature of the 2D glass. For $u>u_c$ one has
$[\sigma](u)=\Sigma_1$. It remains to find the
breakpoint $u_c$, which is determined together with
the effective coupling $\tilde{\mu}$, by the equations
$\sigma(u_c)=2 g e^{- B_{i i}(u_c)/2}$, and
the equation for $\tilde{\mu}$.  $B_{i i+1}$ and the
diagonal correlations $G_{aa}(q,q_z)$ can be computed using
the inversion formula
$G_{aa} = G_c (1 + \int_0^1 du \ [\sigma]/ u^2 (G_c^{-1} + [\sigma]) $
and the above formulae. One finds two equations which
determine $v_c$ and $\tilde{\mu}$:
\begin{equation}
\label{break-result1}
\left( \frac{\tilde{\mu}}{\Lambda}\right)^{1-\tilde{T}}
= \frac{g \tilde{T}}{\Lambda} g(v_c)
\ \ ;\ \ 
\left(\frac{\tilde{\mu}}{\Lambda}\right)^{1-\tilde{T}-\tilde{\Delta}}
= \frac{\mu}{\Lambda} f(v_c)
\end{equation}
where $\Lambda=c q_{max}^2 \sim c (2 \pi/a)^2$.
The following functions have been defined, $g(v) = (1+v)^{(\tilde{T}+1)/2}
(1-v)^{(1-\tilde{T})/2} v^{-1}$ and
$f(v) = g(v) \phi(v)$ with
$\phi(v)=2 v e^{(1+\tilde{T}) \sqrt{(1-v)/(1+v)}-1} /(1+ \sqrt{1-v^2})$.
>From the second equation in (\ref{break-result1})
one recovers the decoupling transition
at $T_c$ for the pure system, and the corresponding value of $\tilde{\mu}$
for $T < T_c$ using that $v_c \to 0$ when $g \to 0$ 
and $f(0)=e^{\tilde{T}}$.

For $\Delta=0$ the equation for $v_c$ is remarkably simple
and reads $g \tilde{T}/\mu = \phi(v_c)$.
In that case, and in the limit of very large cutoff,
the transition from the 3D to the 2D vortex glass RSB solutions is
{\it continuous} and happens when $v_c$=1, i.e $u_c=\tilde{T}$.
The transition line is thus determined by the equation:
\begin{equation} \label{transition0}
\mu = \frac{e \tilde{T}}{2}~g
\end{equation}
Indeed when $v_c \to 1$ one recovers smoothly
the one step solution for the decoupled 2D glass obtained
in Ref. \cite{giamarchi_vortex_global} with $u_c=T/T_c$.

Several effects (such as a finite cutoff, a small
non zero $\Delta$) transform this second order
transition in a (weakly) first order transition (at weak disorder).
When $\Delta > 0$ the above equation becomes:
\begin{equation}
\label{break-result2}
\phi (v_c) \left(g(v_c) \right)^{ \delta }
= {\Lambda \over \mu}
\left( {g \tilde{T} \over \Lambda }\right)^{1- \delta }
\end{equation}
with $\delta= \Delta T/(T_c-T)$.

The function in the left hand side reaches its
maximum before $v_c=1$ and thus upon increasing $T$, the RSB solution ceases 
to exist {\it before} the point $v_c=1$ is reached. Thus there is
now a first order jump to the 2D decoupled phase. The value of
$v_c$ at the jump is given by $v_c^* = (1-\delta^2)/(1+\delta^2)$.
The equation of the transition line is then:
\begin{equation}
{\Lambda \over \mu}
\left( {g \tilde{T} \over \Lambda }\right)^{1- \delta } = 2 \frac{1-\delta}{1+\delta}
\left( \frac{2 \delta^{1-\tilde{T}}}{1-\delta^2} \right)^{\delta} e^{\delta(1+\tilde{T})-1}
\end{equation}
The value of $\tilde{\mu}$ at the transition, i.e the
size of the jump, is given by:
\begin{equation}
\left(\frac{\tilde{\mu}^*}{\Lambda}\right)^{(1-\tilde{T})(1-\delta)}=
\frac{\mu}{\Lambda} \frac{4 \delta^{1-\tilde{T}} }{(1+\delta)^2}
e^{\delta(1+\tilde{T})-1}
\end{equation}

The above result (\ref{transition0}) for the transition from the 
2D to the 3D glass can be recovered in a simple
way by considering the characteristic lengths of the problem.
In the 3D phase (at relatively large $\mu$) one can expand
the interplane coupling terms and obtain a manifold with
bare elastic coefficients: $c_{11} = c (2 \pi/a)^2/l$ isotropically in plane
and $c_z=\mu l (2 \pi/a)^2$ along z. The 3D
translational order correlation length $R_a$ along $z$
was computed in \cite{giamarchi_vortex_global} as
$R_{a,z}^{3d} = a^4 c_{11} c_z/\pi^3 \Delta$ with here $\Delta=2 T g/l$.
One then finds that the above transition corresponds
to the ratio $R_{a,z}^{3d}/l$ being {\it a fixed number}
$R_{a,z}^{3d}/l \approx e=2.7..$.
This can be viewed as a ''Lindemann criterion'' and is in reasonable agreement 
with the one obtained in \cite{nattermann_kierfeld} by a different variational method.
The physical interpretation is thus that the
transition happens when the 3D translational order 
correlation length becomes of order the interplane spacing.
Note that in general it should be the translational order
length $R_a$ which controls the transition and should enter in
a Lindemann type criterion, and {\it not} the Larkin length $R_c$.
Although for a single cosine model, valid at higher temperature
or $\xi \sim a$, one has $R_a \sim R_c$, when one treats model (1) keeping
all harmonics in order to describe the low temperature region,
one finds indeed that it is $R_a$ which is the relevant length.

To interpret further the above results in a simple way one can notice than in
the equations for $\tilde \mu$, relevant $q_z$ momenta are the one close
to the zone boundary ($q_z \sim \pm \pi$). The correlation function
appearing in the equations for $\tilde \mu$ are therefore 
analogous to two dimensional ones, but for the presence of a mass term of
order $\tilde \mu$ in the denominator. One can therefore crudely
replace the equation for $\tilde \mu$ by
\begin{equation} \label{simple}
\tilde\mu = \mu e^{-\frac12 B_{\text{2D}}(r = 1/\sqrt{\tilde \mu})}
\end{equation}
where $B_{2D}(r)$ is the correlation in the purely two dimensional
(disordered) system. On this expression it becomes clear that if
correlation functions in 2D decay too rapidly as a function of distance
there will be no solution of (\ref{simple}) for small $\tilde \mu$, and
the transition will be first order. To get a continuous transition one
need that the right hand side of (\ref{simple}) goes to zero with $\tilde
\mu$, at worst with a finite slope.
This imposes at worst an algebraic decay of the correlation functions
with an exponent smaller than one. This is the case for the pure system
below $T_c$, and for the disordered system with infinite cutoff and no
$\Delta$, where the exponent is frozen at the $T_c$ value (i.e. one). In
the presence of $\Delta > 0$, or if one takes solutions given by
the Cardy-Ostlund RG, for which correlation functions decay as $e^{-\ln(r)^2}$,
one gets a discontinuous transition, as confirmed below. A similar discontinuous
transition was found from the effect of a finite cutoff in
\cite{nattermann_kierfeld}.
Based on (\ref{simple}) one can construct an 
argument from the
2D translational order correlation length for a single plane. 
It was estimated
in \cite{giamarchi_vortex_global} as 
$R_a^{2d} = a (2 \pi \Lambda/T g)^{1/2(1-T/T_c)}$.
This length should be compared with another 2D length 
for the {\it pure} system, i.e the scale below which the coupled 
system is still 2D, $R_{2d} = 2 \pi \sqrt{c/\tilde{\mu}}
= a (\mu e^{\tilde{T}} /\Lambda)^{1/2(1-T/T_c)}$. Remarkably
the transition again occurs when these two lengths become
comparable.

\begin{figure}[htb]
\label{diagphas}
\centerline{ \fig{6cm}{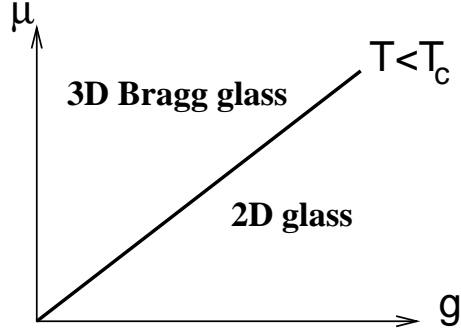} }
\caption{Schematic phase diagram for two coupled
planes in presence of disorder}
\end{figure}

We now turn to the Renormalization Group (RG) method, which can be 
applied to this model near the decoupling transition of the pure system, 
i.e near $T=T_c=4 \pi$, for small disorder $g$ and couplings between planes.
We study the model (\ref{2DXY}). Near $T_c$ one needs to keep only
the lowest harmonic $p=1$, higher harmonics being irrelevant. We have
obtained the (replica symmetric) RG recursion relations for $K_{ij}$, 
$\Delta_{ij}$, 
$g_{ij}$ and $\mu_{ij}$ using Coulomb gas techniques and fermion
methods. This is a generalization of the Cardy Ostlund recursion relations
\cite{cardy_desordre_rg} 
to a set of coupled planes. For simplicity we give here only the 
results for two planes \cite{footnote1}.
The results for $N$ planes will be presented in
\cite{carpentier_decoupling_long} but our preliminary studies show that they
are qualitatively similar (see below). One defines, for two planes:
$K^{-1}_{ij} = (T/T_c) \delta_{ij} + k_{ij}$, $k_c=k_{11}+k_{12}$,
$k=k_{11}-k_{12}$, $\mu = \mu_{12} + g_{12}$, $g=g_{11}$ and 
$\delta=\Delta_{11}-\Delta_{12}$.
One finds the following RG flow equations upon a change of
cutoff $a \to a~e^{l}$ \cite{carpentier_decoupling_long}:
\begin{eqnarray} \label{rg1}
{d \over dl} k & = & \frac{1}{2} ( \mu^{2} -g_{12}^{2}) ~~~ ; ~~~ 
{ d \over dl} \delta  =  \frac{1}{4} ( g^{2} - g_{12}^{2} ) \\ \label{rg2}
{ d \over dl} g &  = &  ( k + k_c )g +
\mu g_{12} - g_{12}^{2} - g^{2} \\ \label{rg3}
{ d \over dl} g_{12} &  = & (- 2 \delta +
k + k_c )g_{12} + \mu g - 2 g_{12}g \\ \label{rg4}
{d \over dl} \mu &  = & 2  ( - \delta + k ) \mu - g_{12} g
\end{eqnarray}
and ${d \over dl } k_c = 0$, ${d \over dl } \Delta_{11} = g^2/4$.
These equations are valid to second
order in all coupling constants and in deviations from $T_c$.
The parameter $k_c=(T_c-T)/T_c$ is the true reduced temperature and
is {\it not} renormalized. $k$ controls the stiffness associated to the
phase difference between the two planes, and is renormalized,
indicating whether the planes are coupled or not.

The previously known limiting cases are the following.
In the absence of disorder, i.e $g=g_{12}=\delta=0$,
one simply recovers a sine-Gordon
model for the relative phase $\phi_1 - \phi_2$: there is a
usual Kosterlitz-Thouless transition, with a separatrix $k=-\mu/2$
between a coupled phase $\mu(l) \to \infty$ and a decoupled 
high-temperature phase ($\mu(l) \to 0$).
In the absence of coupling between planes,
i.e $\mu=0,g_{12}=0,\Delta_{12}=0,k=0$, there is a
a 2D glass phase for $k_c>0$ - characterized 
by the Cardy-Ostlund attractive fixed point $g^*=k_c$ 
and $\delta(l) \to \infty$ - as well as a high temperature phase 
for $k_c<0$, where disorder is
irrelevant and $g(l) \to 0$ and $\delta(l) \to \delta^*$.

In presence of both disorder and interplane coupling, 
we find, via
numerical solution of the equations that there are {\it three}
very different regimes of the flow with marked separatrices between them 
(we start from 
the initial condition $k=k_c, g_{12}=0, \Delta_{ij}=0$
corresponding to model (4)). There is a decoupled 2D glass 
regime ($\mu(l) \to 0, g_{12}(l) \to 0, \delta \to \infty, k(l) \to k^*$),
a 3D coupled glass phase (all constants become large) and a high
temperature phase ($\mu(l),g_{12}(l),g(l) \to 0$, 
$\delta(l) \to \delta^*, k(l) \to k^*$). These regimes are very
likely to correspond to three different phases, the uncertainty
on the exact position of the separatrices being
quite small \cite{footnote}. We thus arrive at the phase diagram of Fig. 1. 
In particular the phase transition between the 2D and 3D glass phases occurs
approximately for:
\begin{equation} \label{transition}
\mu \approx  C~~g  
\end{equation}
with $C \approx 0.5-0.6$, and only a weak dependence on temperature. 
This result is in good agreement with the result
for the transition line from the GVM. One also notices that,
when approaching the 2D-3D glass separatrix from the 2D glass phase,
the interplane coupling always becomes vanishingly small at a {\it fixed} characteristic
length $l^*(g)$, which diverges as $g \to 0$ but remains fixed close to the
transition. We find numerically that $l^*(g) \sim 1/(2 k_c) \log(1/g)$
for small $g$ and thus we identify this length with the 2D Larkin length.
This may be interpreted as the 2D to 3D glass transition being 
first-order, also 
in agreement with the GVM.
Finally, the study for N planes gives similar results
for the phases. The separatrix remains roughly of the form 
(\ref{transition}), with $C \approx 0.7-1.0$, but
deviations from the linear
behaviour is observed \cite{carpentier_decoupling_long}
at small $g$ and near $T_c$.

In conclusion we have shown that in a layered geometry, the Bragg 
glass phase is stable at weak disorder. Upon increase of disorder
it undergoes a transition into a topologically disordered glass.
In this geometry this glass is in fact a decoupled 2D glass.
The transition is found to occur roughly when the translational
order length $R_a$ is of the order of the lattice spacing $a$ 
as suggested in \cite{giamarchi_vortex_global}.
This model has some peculiarities that make generalization to
real vortex lattices difficult. However, the fact that it confirms
the existence of a stable Bragg glass is an additional indication, 
besides the energy arguments of \cite{giamarchi_vortex_global}, 
that the Bragg glass occurs in real vortex lattices. 
It also confirms that the criterion $R_a \gtrsim a$ may indeed be used 
as stability criterion for the Bragg glass. Our conclusions
agree with the one of \cite{nattermann_kierfeld}.

We are grateful to T. Nattermann for interesting
discussions which motivated this work.

%\end{references}


\begin{thebibliography}{99}
\bibitem{footnote1} After this work was completed we learned that
this model for two planes was studied by 
a RG analysis in J. Kierfeld, Diploma thesis,
Universit\"at zu K\"oln (1993);
\bibitem{footnote} initial conditions which ultimately flow to
the 2D glass fixed point tend to leave, at intermediate scales, 
the perturbative
region where (\ref{rg1},\ref{rg2},\ref{rg3},\ref{rg4}) are valid. This results
lead to an uncertainty in locating the separatrix.
\end{thebibliography}
\end{document}